\documentclass[namedreferences,hyperref,optionalrh]{spr-sola}

\usepackage{graphicx}        
\usepackage{tablefootnote}
\usepackage{float}
\usepackage{wrapfig}
\usepackage{url}  

\usepackage{cancel}
\usepackage{ulem}

\usepackage{amsmath}
\usepackage{parskip}

\usepackage{hyperref}

\newcommand{\aap}{{\it Astron. Astrophys.}}

\newcommand{\apjl}{{\it Astrophys. J. Lett.}}
\newcommand{\apjs}{{\it Astrophys. J. Supl.}}

\begin{document}

\begin{frontmatter}
\title{Emission Measures Demystified}




\graphicspath{{./}{figures/}}

\author[addressref={aff1},email={brian.r.dennis@nasa.gov}]{\inits{B.R.}\fnm{Brian~R.}~\snm{Dennis}\orcid{0000-0001-8585-2349}}



\author[addressref={aff2}]{\inits{K.J.H.P.}\fnm{Kenneth}~J.~H.~\snm{Phillips}}






\address[id=aff1]{Solar Physics Laboratory, Code 671, NASA Goddard Space Flight Center, 
Greenbelt, MD 20771, USA}

\address[id=aff2]{Scientific Associate, Earth Sciences Dept., Natural History Museum, Cromwell Road, London SW7 5BD, UK}


\begin{abstract}

We review the terms, spectral radiance and spectral irradiance, and show how their precise definitions are crucial for interpreting observations made with different instruments covering widely different energy or wavelength ranges. We show how the use of column and volume emission measures in different solar physics and astrophysics communities has led to confusion in relating measured extreme ultraviolet and soft X-ray spectra with theoretical spectra generated, in particular, using CHIANTI. We describe a method for obtaining spatially integrated X-ray line and continuum spectra using CHIANTI that requires a column emission measure when only the plasma temperature and volume emission measure, but not the source area, are known from observations.

\end{abstract}


\end{frontmatter}

\section{Introduction}
\label{sec:intro}

Much of the solar ultraviolet (UV), extreme ultraviolet (EUV), soft X-ray (SXR), and hard X-ray (HXR) emission from the quiet and active solar corona and solar-flare  plasmas is optically thin. Excitation of both line and continuum emission is by collisions of free electrons with target protons and heavier ions. As a result, the amount of emission per unit volume depends on the number density of both ions $N_i$ and electrons $N_e$, together with temperature-dependent functions that are Maxwellian averages of the cross sections of the emission process for a particular electron temperature $T_e$. 

Line emission has been discussed by \cite{2012uxss.book.....P,2018LRSP...15....5D}, 
and continuum emission (free--free or bremsstrahlung)
by \cite{1982ApJ...260..875D}. In addition, free--bound emission can have comparable intensity over certain temperature and energy ranges.

We take the emitting plasma to be located in the solar corona and to be optically thin, assumptions that are valid for all but a few resonance lines for which resonant scattering may be significant. Excitation by radiation, including fluorescence, is also not considered here. We also assume a steady state, or at least only very slowly varying conditions, for the observed plasma, appropriate for the quiet-Sun corona, non-flaring active regions, and slowly evolving flares


The CHIANTI code (Appendix \ref{sec:CHIANTI} and \cite{2021ApJ...909...38D}) requires as input a column emission measure in order to compute a spectrum. In contrast, the OSPEX set of routines (Appendix \ref{sec:OSPEX}), widely used for X-ray observations made by the {\it Reuven Ramaty High-Energy Solar Spectroscopic Imager} (RHESSI, \cite{2002SoPh..210....3L}) and the Spectrometer Telescope for Imaging X-rays (STIX, \cite{2020A&A...642A..15KKruckerHurfordGrimmA&A2020}) on the {\it Solar Orbiter} spacecraft, outputs temperature and volume emission measure for the thermal spectra of solar flares. 

This Research Note provides a method for comparing measured EUV, SXR, and HXR solar spectra with theoretical spectra calculated from atomic codes. We define the terms radiance and irradiance used to characterize the emissions from a thermal plasma. We then define column and volume emission measures, and the relation between the two. 

\section{Radiance vs. Irradiance}
\label{Int_vs_Irrad}

We define radiance and irradiance, and the equivalent spectral radiance and spectral irradiance, following the Oxford Dictionary of Physics \citep{2009ODP...2009dictionary} and Wikipedia.\footnote{https://en.wikipedia.org/wiki/Radiance and https://en.wikipedia.org/wiki/Irradiance} 

\begin{figure}
\begin{center}
\includegraphics*[width=0.7\textwidth,  
    angle = 0, trim = 0 0 0 0]
    {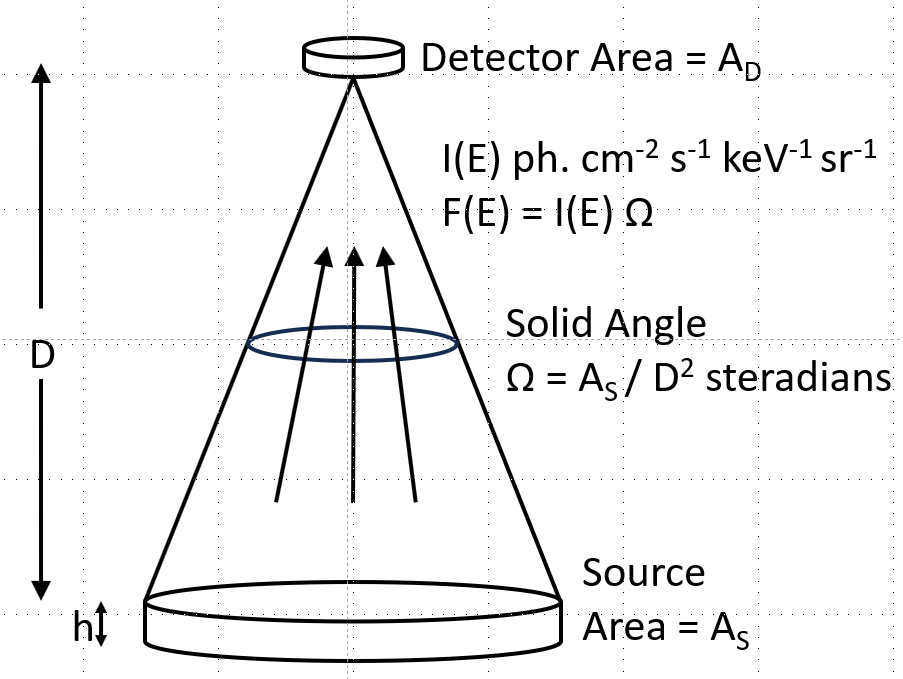}
\end{center} 
\caption{Illustrating the radiance and irradiance of the emission measured by a detector at a distance $D$ from a cylindrical source with area $A_S$ and thickness $h$.}
\label{fig:IntensityFlux}
\end{figure}

\textit{Irradiance} ($F$) is the radiant flux density from a source that is \textbf{\textit{received}} at a remote surface. $F$ is the number of photons or ergs reaching the surface per unit (detector) area and time. Spectral irradiance is the irradiance per spectral interval, with symbol $F_\lambda$ for wavelength (e.g. \AA), $F_\nu$ for frequency (e.g. Hz), and $F_E$ for photon energy (e.g. keV). This is illustrated in Figure~\ref{fig:IntensityFlux}, where the source is a cylinder with thickness, $h$, and area, $A_S$. Photons are emitted isotropically but only those in the indicated solid angle are received by the detector at a distance $D$ from the source. The photon spectral irradiance received by the detector is, in energy units, $F_E$~photons~cm$^{-2}$~s$^{-1}$~keV$^{-1}$. Note that the unit, cm$^{-2}$, pertains to the area of the detector, not the area of the source.\footnote{Following most usage in solar physics, c.g.s. units are used throughout this paper.} 

The total emission from the source assuming isotropy is $4\pi{D^2}F$ photons s$^{-1}$ keV$^{-1}$, where $4\pi{D^2}$ is the surface area of the sphere with a radius of $D$ centered on the source.  The spectral irradiance from a source is measured by a detector such as RHESSI and can be computed from the observations using the OSPEX routines (Appendix~\ref{sec:OSPEX}).

\textbf{Radiance ($I$)} is the radiant flux \textbf{\textit{received}} from a source per unit area, time, and solid angle. Thus, it is the same as the irradiance ($F$) but per unit solid angle into which the flux is emitted. Consequently, it is independent of the distance of the receiving surface from the source. This is the reason why it is generated by CHIANTI. Spectral radiance is the radiance per spectral interval, with symbol $I_\lambda$ for wavelength (e.g. \AA), $I_\nu$ for frequency (e.g. Hz), and $I_E$ for photon energy (e.g. keV).

The total emission from the source assuming isotropy is $4{\pi}I$ with units of photons cm$^{-2}$~s$^{-1}$~keV$^{-1}$, where $4\pi$ is the number of steradians of the sphere centered on the source.

\begin{figure}
\begin{center}
\includegraphics*[width=0.35\textwidth,  
    angle = 0, trim = 0 0 0 0]
    {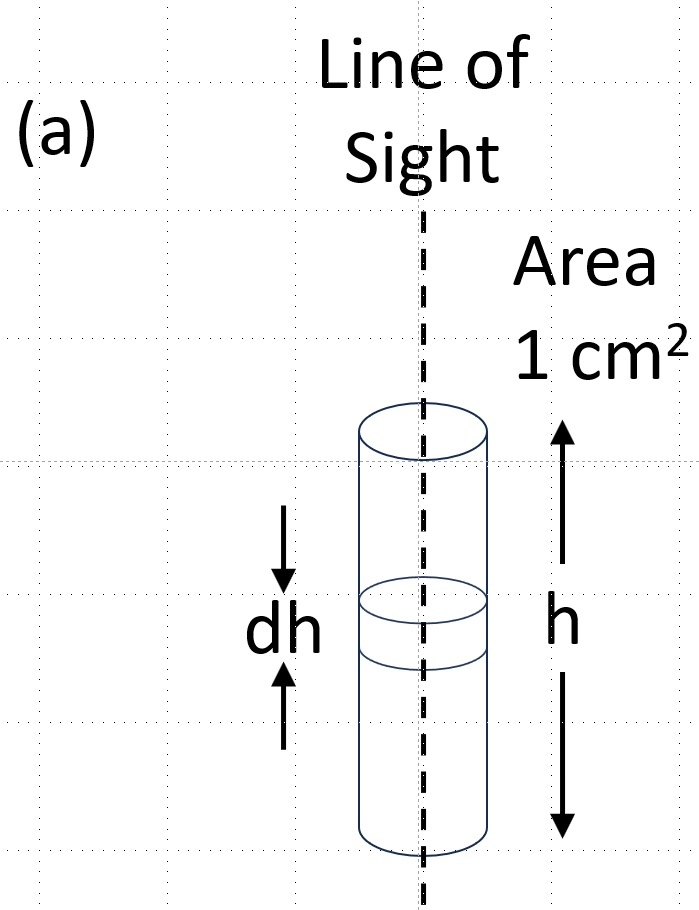}
\includegraphics*[width=0.55\textwidth,  
    angle = 0, trim = 0 0 0 0]
    {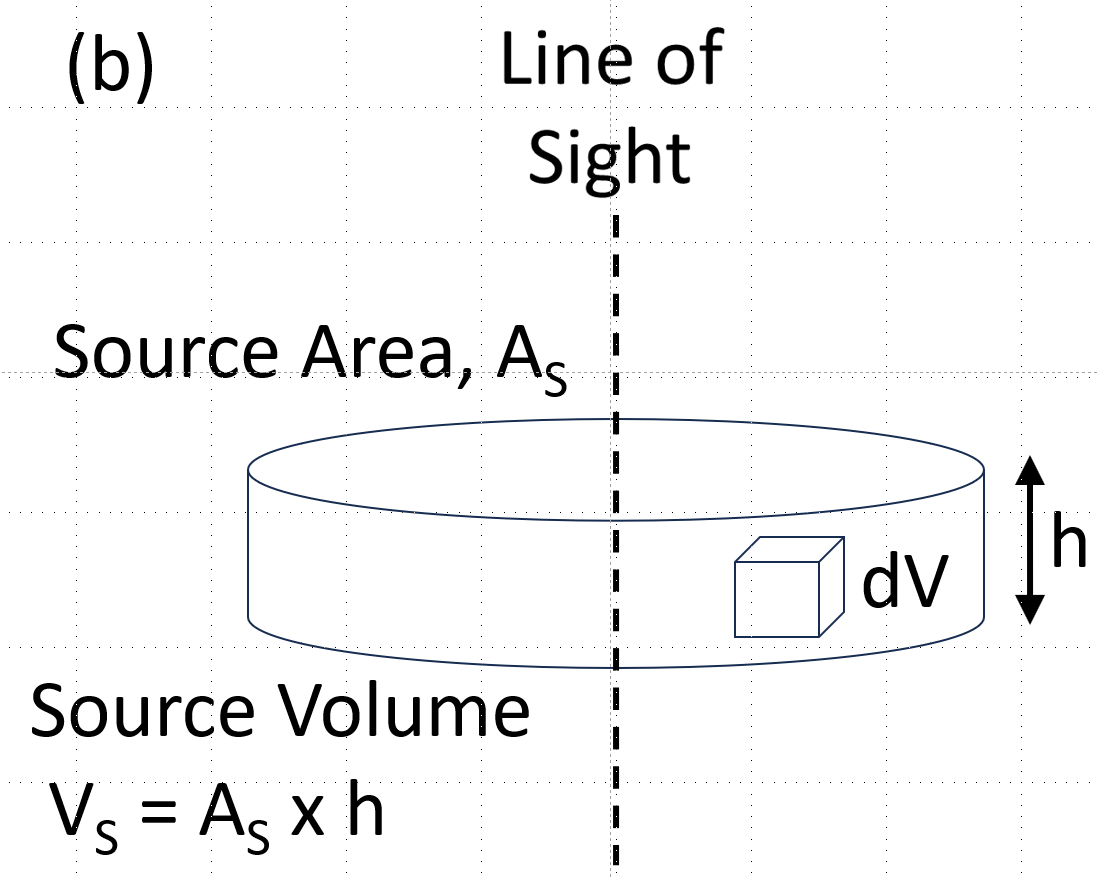}
\end{center} 
\caption{Illustrating the difference between column and volume emission measures. (a) The column emission measure is a measure of the emission from a cylindrical source along the line of sight with unit (1~cm$^2$) cross-sectional area. (b) The volume emission measure is a measure of the emission from the whole volume of a source that in the simplest case shown here is a cylinder with a cross-sectional area, A$_S$, and thickness, $h$. This leads to the relation given by Equation~\ref{eq:EMhA}. }
\label{fig:EMs}
\end{figure}

The solid angle into which the detected photons are emitted is the solid angle, $\Omega$, subtended by the source at the detector, where

\begin{equation}\label{eq:Omega}
    \Omega~=~A_S~/~D^2.
\end{equation}

The relation between $I$ and $F$, again assuming isotropy is just the solid angle $\Omega$ subtended by the source with area $A_S$ at the detector, i.e.

\begin{equation}\label{eq:FIOmega}
    F = I\Omega = I A_S/D^2.
\end{equation}


\section{Emission Measures}\label{EMs}

Within the limitations identified above, the amount of both line and continuum emission can be characterized by the same emission measures that depend only on the product of the densities of the free electrons ($N_e$) and target ions ($N_i$) integrated over the source volume. $N_i$ is related to $N_e$ by the abundance of the ions, which equals the abundance of the emitting element and a function of temperature which defines the ion fraction through an ionization equilibrium calculation. Thus, the emission measure, which is a property of the emitting plasma, is the integral of $N_e^2$ over a height range or volume. This applies to continuum emission (free--free or bremsstrahlung, free--bound, and two photon) as well as line emission apart from a small minority of strong resonance lines for which resonant scattering is important. The emission measure is undefined for optically thick plasma. 

Two different forms of emission measure are defined -- the column emission measure and the volume emission measure. They are illustrated in Figure~\ref{fig:EMs}.

\subsection{Column Emission Measure}
\label{CEM}

The emission of a solar source as measured along the line of sight from a detector is illustrated in Figure~\ref{fig:EMs}(a). This gives rise to the concept of a \textbf{column emission measure, $EM_h$}, defined by  

\begin{equation}\label{eq:EMh}
    EM_h = \int_h N_e^2~dh~{\rm cm}^{-5}
\end{equation}

\noindent where $N_e$ is the number density of free electrons (in cgs units cm$^{-3}$). This definition was given by \cite{1964SSRv....3..816PPottaschSSRv1964} and is used for both UV and EUV emission, and by the CHIANTI suite of programs. It can be approximated by $N_e^2~h$ if the emitting region is isothermal and homogeneous along the line of sight. Note $EM_h$ should not be confused with the amount of material in the column through the source along the line of sight. That would be $(N_i~+~N_e)~\times~A_S~\times~h$.
 
\subsection{Volume Emission Measure}

When the emission from a large or unresolved source is measured at a given distance, then it is common to use the concept of a  \textbf{volume emission measure, $EM_V$}, as illustrated in Figure~\ref{fig:EMs}(b). It is defined by

\begin{equation}\label{eq:EMv}
    EM_V = \int_V N_e^2~dV~{\rm cm}^{-3}
\end{equation}

It is generally used for X-ray observations, such as from GOES and RHESSI, where the source dimensions are poorly defined or are unknown. It may be computed from the X-ray spectral observations using OSPEX (Appendix~\ref{sec:OSPEX}) or similar data analysis packages.

\section{Why the Volume Emission Measure is \texorpdfstring{$N_e^2 V$}{Ne^2 V} }
\texorpdfstring{\hfil}{\space}
\label{Why N2}

 \begin{figure}[htbp]
\begin{center}
\includegraphics*[width=0.6\textwidth,  
    angle = 0, trim = 0 0 0 0]
    {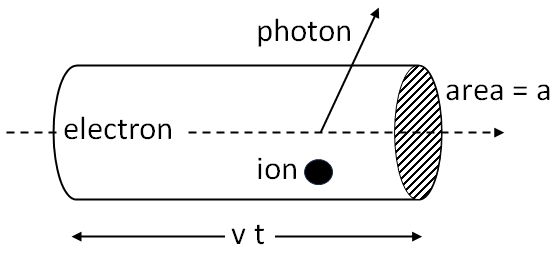}
\end{center} 
\caption{Diagram showing that, on average, one photon will be emitted in time, t, when one ion lies within the volume defined by the distance traveled by an electron with velocity, v, and the cross-sectional area, a, of the interaction that produces the photon.}
\label{fig:OnePhoton}
\end{figure}

The concept of an emission measure arises from considerations of the rate of photon-producing encounters between free electrons in the source and target ions (e.g. \cite{2004psci.book.....A}(p.~44)). The fact that the emission intensity is proportional to the product, $N_e~N_i$, can be appreciated using the diagram in Figure~\ref{fig:OnePhoton} showing the path of a single electron traveling with velocity, $v$, through the source. The electron will travel a distance of $v~\times~t$ in a time, $t$. A single photon will be produced if just one ion is within the indicated cylindrical volume with a length of $v~\times~t$ and an area equal to the cross-sectional area, $a$, of the particular interaction involved, either free--free or free--bound for continuum emission or excitation for line emission. On average, the number of ions in this volume is  $N_i~a~v~t$. Setting this number to one shows that

\begin{equation}
    t = (N_i~a~v)^{-1} .  
\end{equation}

Thus, each electron in the source will produce one photon in time, t, and the rate of photons emitted from an element of the source volume, $dV$, will be just the total number of electrons in this volume ($N_e~dV$) divided by the time for each electron to produce one photon, i.e.

 \begin{equation}
    N_e~dV~/~t~=~N_e~N_i~a~v~dV  
 \end{equation}
 
 \noindent Integrating this over the whole volume $V$, then gives the total rate of photons emitted by the source 

 \begin{equation}
     photon~rate = \int_VN_e~N_i~a~v~dV.
 \end{equation}
 
\noindent Assuming that $a$ and $v$ are independent of the position in the source, they can be removed from the integral over volume. Also, for an electrically neutral plasma, $N_i \approx N_e$ leaving just $\int_V N_e^2~dV$ for the definition of the volume emission measure.

\section{Relation between column and volume emission measures}
\label{sec:Relation}

As shown in Figure~\ref{fig:EMs}(b), the emitting region's volume $V$ can be taken to be $A h$ where $A$ is the area of the emitting region perpendicular to the line-of-sight and $h$ is the thickness. Thus, the volume emission measure, $EM_V$, is just the column emission measure, $EM_h$, multiplied by the source area $A$ as seen from the detector, i.e.,

\begin{equation}\label{eq:EMhA}
   EM_V = EM_h \times A. 
\end{equation}

\noindent The source area subtends a solid angle $\Omega~=~A/D^2$ at the detector receiving the radiation which is at a distance $D$ from the source. 
In that case, the spectral irradiance $F_\lambda$ and spectral radiance $I_\lambda$ are related by $F_\lambda~=~I_\lambda~A~/~D^2$ (Eq.~\ref{eq:FIOmega}).
Then, using Eq.~\ref{eq:FIOmega} and Eq.~\ref{eq:EMhA}, the spectral irradiance $F_\lambda$ per unit volume emission measure is related to the spectral radiance $I_\lambda$ per unit column emission measure by

\begin{equation}\label{eq:F_I_relation}
\boxed{\frac{F_\lambda}{EM_V} = \frac{I_\lambda~A}{EM_h~A} \times \frac{1}{D^2} = \frac{I_\lambda}{EM_h} \times \frac{1}{D^2}},
\end{equation}

\noindent or, expressed in words, the spectral irradiance per unit volume emission measure equals the spectral radiance per column emission measure multiplied by $1/D^2$. It is independent of the emitting region's area $A$. 


Equation~\ref{eq:F_I_relation} gives the procedure for obtaining the spectral irradiance ($F_\lambda$) for a specific volume emission measure given the spectral radiance ($I_\lambda$) obtained from CHIANTI for a specific column emission measure. It requires that we know only the distance, D, between the source and the detector but not the source area or volume.

For Earth-orbiting solar spacecraft such as {\it Skylab}, {\it Solar Maximum Mission}, {\it Yohkoh}, {\it RHESSI}, and {\it Hinode}, $D$ is, on average, 1~A.U.~or $1.496\times 10^{13}$~cm, so $1/D^2 = 4.53\times 10^{-27}$~cm$^{-2}$. Note that $D$ varies by $\pm 1.4$\% throughout the year (largest in January, smallest in July), a point that may be of importance for analysis of long-term spectro-photometry. For {\it SOHO} located at the inner Lagrangian point, $D = 148.5 \times 10^6$~km $= 1.485 \times 10^{13}$~cm  so $1/D^2~=~4.47\times 10^{-27}$~cm$^{-2}$. For Sun-orbiting spacecraft such as {\it STEREO-A} and {\it STEREO-B} and {\it Solar Orbiter}, $D$ varies widely with time so that the instantaneous value at the time of observation should be used in Equation~\ref{eq:F_I_relation}.

\begin{figure}
\centerline{\includegraphics[width=1.0\textwidth,clip=,angle=0]
{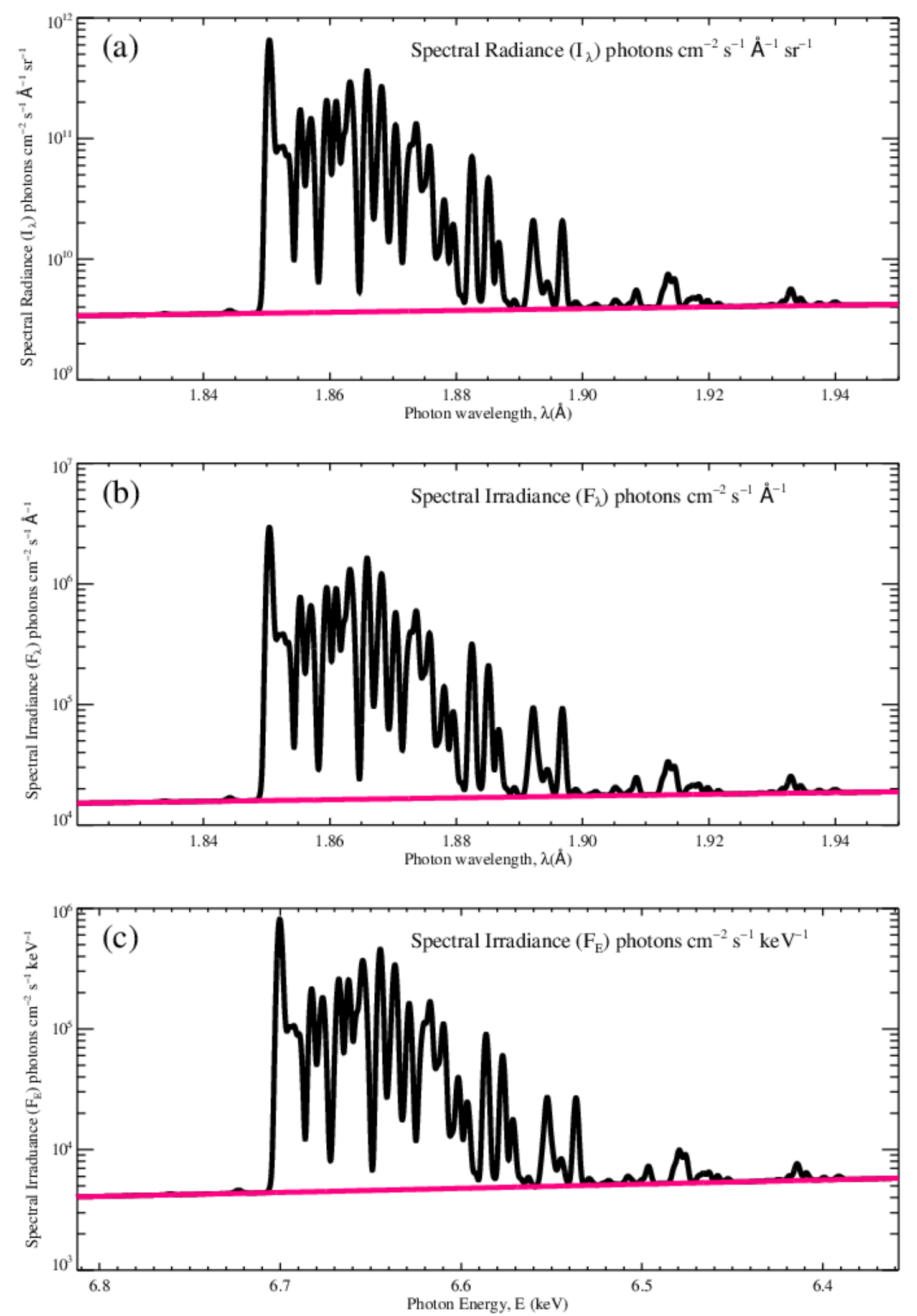}}
\caption{X-ray spectra emitted by a coronal flare plasma showing lines of highly ionized Fe atoms to illustrate Equation~\ref{eq:F_I_relation}. Top panel (a): spectral radiance $I_\lambda$ (photon cm$^{-2}$ s$^{-1}$ \AA$^{-1}$ sr$^{-1}$) as output from CHIANTI (version 10.1) for $T_e = 20$~MK, column emission measure $EM_h = 10^{27}$~cm$^{-5}$. Middle panel (b): spectral irradiance $F_\lambda$ (photon cm$^{-2}$ s$^{-1}$ \AA$^{-1}$) for $EM_V = 10^{48}$~cm$^{-3}$ calculated from spectrum (a) using Equation~\ref{eq:F_I_relation}. Bottom panel (c): spectral irradiance $F_E$ (photon cm$^{-2}$ s$^{-1}$ keV$^{-1}$) in energy (keV) units on the abscissa. The red curve in each case is the spectrum of the continuum emission.} 
\label{fig:Fe_line_sp}
\end{figure}

We illustrate the use of Equation~\ref{eq:F_I_relation} with the three spectra shown in Figure~\ref{fig:Fe_line_sp}.
The first spectrum in Figure~\ref{fig:Fe_line_sp}(a) was generated using the CHIANTI (version 10.1) spectral synthesis routine accessed either from a Graphics User Interface (GUI) initiated in the Interactive Data Language (IDL) with the command ch\_ss or the command-line equivalent (see Appendix \ref{sec:CHIANTI}). It shows the output $I_\lambda$ spectrum in the range 1.83 - 1.95~\AA\ for an isothermal plasma with a typical flare temperature, $T_e = 20\times 10^6$~K~or~log$_{10}T_e$ (K) $ = 7.3$, and the default column emission measure of $10^{27}$~cm$^{-5}$. This wavelength range includes the well-known group of highly ionized Fe lines (Fe~{\sc xxv} lines with satellites) that have been observed both with high-resolution SXR spectrometers like the Bent Crystal Spectrometer (BCS) on the {\it Solar Maximum Mission} \citep{2017SoPh..292...50RRapleySylwesterPhillipsSoPh2017} as well as broad-band detectors such as those on RHESSI \citep{2002SoPh..210....3L}. The wavelength scale is in \AA\ and the ordinate scale is in units of $I_\lambda$, viz. photons~cm$^{-2}$~s$^{-1}$~sr$^{-1}$~\AA$^{-1}$. The wavelength resolution (FWHM) for this spectrum is 0.001\AA, approximately that of the SMM BCS. 

Apart from certain density-sensitive lines, the line and continuum components of the isothermal $I_\lambda$ spectrum shown in Figure~\ref{fig:Fe_line_sp}(a) all scale linearly with both column and volume emission measure. This allows Equation~\ref{eq:F_I_relation} to be used over a wide range of emission measures to transform an $I_\lambda$ spectrum for a given column emission measure into an $F_\lambda$ or an $F_E$ spectrum for an arbitrary volume emission measure. For example, the following relations were used to transform the spectrum shown in Figure~\ref{fig:Fe_line_sp}(a) that is for a column emission measure of $10^{27}$~cm$^{-5}$ into the spectral irradiance $F_\lambda$ shown in Figure~\ref{fig:Fe_line_sp}(b) for a volume emission measure $10^{48}$~cm$^{-3}$ (a typical value for a medium-sized flare):

\begin{description}\label{ItoF}
   \item[ ] \begin{center} $F_\lambda = I_\lambda \times 10^{48}/(10^{27} \times D^2)$
   \item[ ]  or for $D = 1$~A.U. 
    \item[ ] $F_\lambda = 4.468~10^{-6} \times I_\lambda$.
\end{center}
\end{description}

The conversion of the $F_\lambda$ spectrum of Figure~\ref{fig:Fe_line_sp}(b) into the $F_E$ spectrum shown in Figure~\ref{fig:Fe_line_sp}(c) plotted against energy units (keV) must take into account the change in units from \AA\ (Angstrom) to keV in both axes.  This is done using the following relations between photon energy in ergs or keV and frequency ($\nu$) in Hz or wavelength ($\lambda$) in cm or \AA: 
\begin{equation}
    E = h \nu = \frac{h~c}{\lambda}    
\end{equation}
 \noindent where $h$ is Planck's constant = $6.626~\times 10^{-27}~$erg~Hz$^{-1} = 4.136~\times 10^{-15}$~eV~s, and 
 $c$~is the velocity of light $= 3.00 \times 10^{10}$~cm~s$^{-1}$.
Thus, if $\lambda$ is in \AA, then  

\begin{equation}
    E ({\rm keV}) = \frac{12.399}{\lambda(\text{\AA})}~{\rm or} ~E ({\rm erg}) = \frac{2.0~\times 10^{-8}}{\lambda(\text{\AA})}
\end{equation}

Differentiating these relationships gives the following:

\begin{equation}{}
    \frac{dE}{d\lambda} = -\frac{hc}{\lambda^2} 
    = -\frac{E^2}{hc} 
\end{equation}

with units of erg~cm$^{-1}$~or eV~cm$^{-1}$.
We require that the differential irradiance or flux spectrum be independent of the units for $\lambda$, i.e that

\begin{equation}
    F(\lambda) d\lambda = F(E) dE
\end{equation}

\noindent Thus, $F(E) = F(\lambda) d\lambda/dE$

\noindent or  $F(\lambda) \lambda^2/(h c)$ photons cm$^{-2}$ s$^{-1}$ keV$^{-1}$ 

\noindent or  $F(E) = F(\lambda) (h c)/E^2$ photons cm$^{-2}$ s$^{-1}$ keV$^{-1}$  

 So to convert a spectrum of fluxes in photons cm$^{-2}$~s$^{-1}$~\AA$^{-1}$ plotted against the wavelength in \AA~to a spectrum in photons cm$^{-2}$~s$^{-1}$~keV$^{-1}$ plotted against the photon energy in keV, the following adjustments must be made:
\begin{itemize}
  \item Multiply the fluxes by  $\lambda^2 /hc = \lambda^2 /12.399$ where $\lambda$ is in \AA,\\
  or multiply the fluxes by  $12.399/E^2$ where E is the bin energy in keV.
 \item Convert the photon wavelength in \AA~to photon energy in keV using the relation E = $12.399 / \lambda$ keV.  \end{itemize}

The resulting spectrum is shown in Figure~\ref{fig:Fe_line_sp}(c), with the irradiance in units of photons cm$^{-2}$~s$^{-1}$~keV$^{-1}$ and the energy $E$ scale decreasing from left to right for comparison with the plots vs. wavelength. The energy resolution corresponds to the wavelength resolution of Figure~\ref{fig:Fe_line_sp}(a) and (b), viz. 3.4~eV. Incidentally, this compares with 1~keV for RHESSI and $\sim 120$~eV for the XIS instrument on the {\it Suzaku} spacecraft \citep{2007PASJ...59S..23K} and so the spectral resolution used for the figure is much higher than is possible with present-day instrumentation.

\section{Using Volume Emission Measure in CHIANTI}
\label{sec:Fooling}

\begin{figure}
\begin{center}
\includegraphics*[width=0.9\textwidth,  
    angle = 0, trim = 0 0 0 0]
    {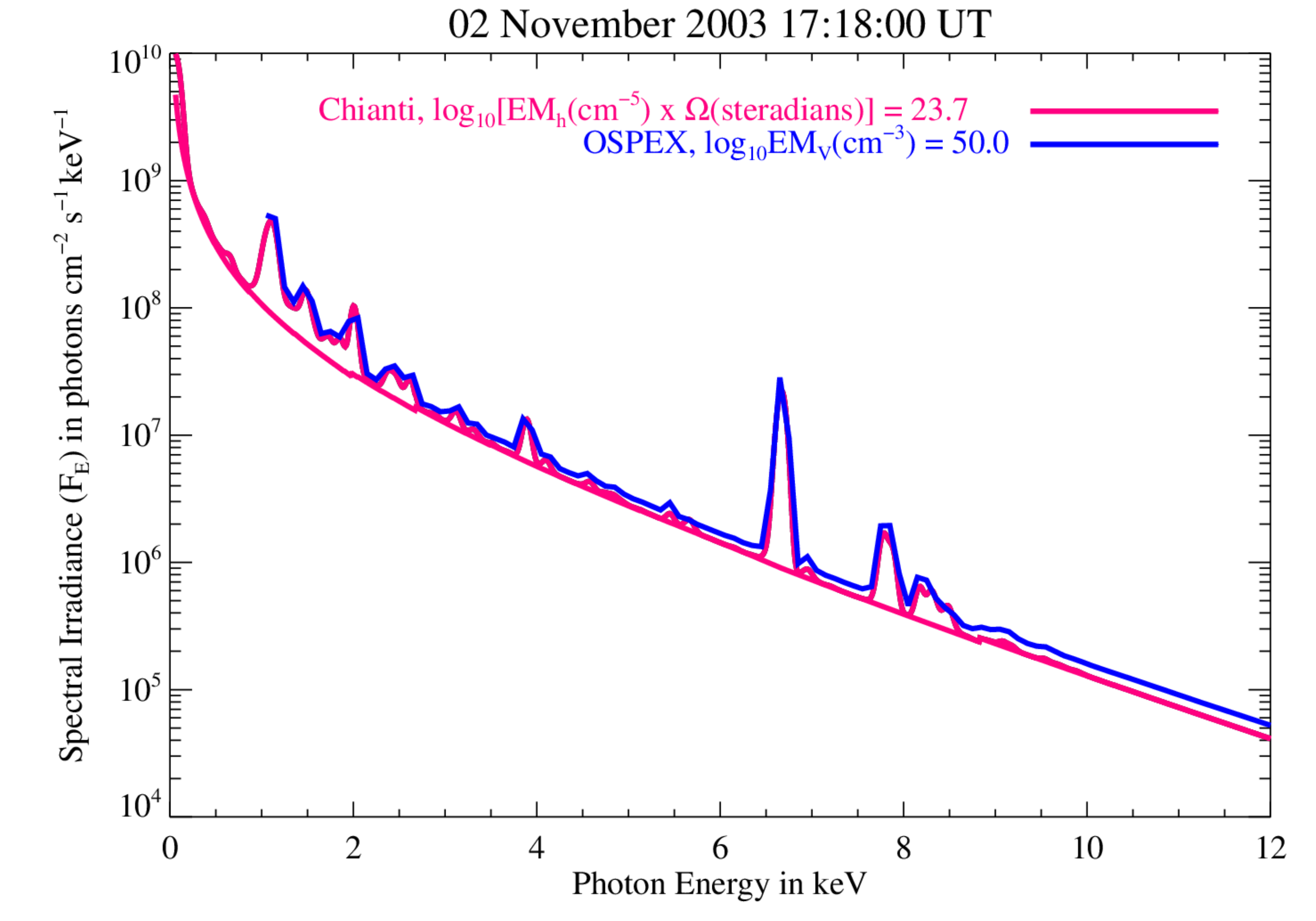}
\end{center} 
\caption{Comparison of X-ray spectra for the flare on 02 Nov.~2003 at 17:18 UT made with OSPEX (blue) and CHIANTI (red). The two red curves were made using the CHIANTI GUI (ch\_ss) with the input \textbf{effective} column emission measure calculated using Equation~\ref{eq:log10EMV/DV^2} (see text). The smooth red curve is for the continuum and the other red curve includes the line emission broadened with an instrumental FWHM of 0.1~keV.}
\label{fig:spectra}
\end{figure}

Frequently, we want to use CHIANTI to generate a photon spectrum when we know only the temperature and the volume emission measure.  As an example, we used the IDL GOES workbench\footnote{https://hesperia.gsfc.nasa.gov/goes/goes.html} to determine the following values for the flare on 02 Nov.~2003 at 17:18 UT: log$_{10}T({\rm K}) = 7.4$ and $EM_V = 10^{50}$~cm$^{-3}$. We used the OSPEX ``vth'' function to plot the spectral
irradiance, $F_E$, vs. energy in keV shown in blue in Figure~\ref{fig:spectra} extending down to 1 keV, the OSPEX limit. The spectrum shown in red is from CHIANTI. Both spectra are for an isothermal source with the same plasma temperature, log$_{10}T(K) = 7.4$ and for coronal abundances. 

The CHIANTI spectrum can be obtained in two different ways - (1) using Equation~\ref{eq:F_I_relation} to convert from $I_E$ to $F_E$ as described in Section~\ref{sec:Relation} or (2) using an ``effective'' column emission measure entered into the CHIANTI GUI that produces the correct $F_E$ spectrum.

(1) The first method is a numerical procedure based on Eq.~\ref{eq:F_I_relation}. 
It starts with the CHIANTI-generated radiant intensity $I_E$ spectrum for a default column emission measure $EM_h$ of $10^{27}$~cm$^{-5}$. Since GOES is in Earth orbit, \hbox{$D=1$~AU} to within $\pm 1.4$\%.  Then, for $EM_V= 10^{50}$~cm$^{-3}$, we have that $F_E = I_E \times (10^{50}/10^{27}) / (1/1.495 \times 10^{13} = 4.475 \times 10^{-5} I_E$. This is plotted in red in Figure~\ref{fig:spectra}. It extends from 12~keV down to below 0.1 keV,  and includes separate line and continuum components.

(2) An alternate way of computing the $F_E$ spectrum using CHIANTI is to determine a value to enter as an ``effective'' column emission measure that includes a solid angle. Since the $F_E$ spectrum is just the $I_E$ spectrum times this solid angle, the resulted spectrum generated by CHIANTI is then the $F_E$ spectrum, even though it is still labeled with units of photons cm$^{-2}$~s$^{-1}$~keV$^{-1}$~sr$^{-1}$. 

The correct value of the ``effective'' column emission measure can be determined by considering the simple case shown in Figure~\ref{fig:IntensityFlux}. The source is a uniform cylinder aligned perpendicular to the line of sight from the observer with thickness, $h$, and front surface area, 
$A({\rm cm}^2$). In that case, the volume of the source, $V = A~\times~h$, and 

\begin{equation}\label{eq:EMhXA}
EM_V = EM_h~\times~A~{\rm cm}^{-3}    
\end{equation}

However, we generally do not know the area of the source, especially if we are computing the spatially integrated spectrum with OSPEX or using the GOES X-ray data. Furthermore, we want to calculate the total irradiance from the source at the detector, not the irradiance per unit sold angle. We can exploit the fact that we do know the distance, $D$, from the source to the detector.  The solid angle $\Omega$ subtended by a solar source with area $A$ at a distance $D$ away from the source as given by Eq.~\ref{eq:Omega}. Then, substituting for $A$ in Eq.~\ref{eq:EMhXA} gives

\begin{equation}\label{eq:EMhXOmega}
EM_V = EM_{h}~\Omega~D^2~{\rm cm}^{-3}~\text{or}
\end{equation}

\begin{equation}\label{eq:EMV/D^2}
EM_h~\times~\Omega = {\frac{EM_V}{D^2}}~{\rm cm}^{-5}~{\rm}sr~\text{or}
\end{equation}

\begin{equation}\label{eq:log10EMV/DV^2}
{\rm log}_{10}(EM_h~\times~\Omega) = {\rm log}_{10}EM_V - {\rm log}_{10}D^2
\end{equation}

Since GOES is in Earth orbit, D~=~1~astronomical unit~(AU) on average, i.e.

\begin{description}
\item[ ] $D~=~1.496 \times 10^{13}~{\rm cm}$, 
$D^2~=~2.238 \times 10^{26}~{\rm cm}^2$, and 
${\rm log}_{10}D^2({\rm cm}^2)~=~26.35$.
\end{description}

\noindent Thus, with the measured value of $EM_V = 10^{50}~{\rm cm}^{-3}$, if we enter the value of 50~-~26.35~=~23.65 into CHIANTI instead of the usual $log_{10}EM_h({\rm cm}^{-5})$, then the spectrum generated will be the spectral radiance times a source solid angle, i.e. the spectral irradiance. It will therefore have units of photons~cm$^{-2}$~s$^{-1}$~keV$^{-1}$ instead of the usual photons~cm$^{-2}$~s$^{-1}$~sr$^{-1}$~keV$^{-1}$, i.e. without the ``${\rm sr}^{-1}$.'' The current version of CHIANTI still labels the spectrum with ``${\rm sr}^{-1}$'' in the units when this technique is used. The result can be validated by comparing in Figure \ref{fig:spectra} the resulting spectrum with that produced by OSPEX using the ``vth'' function with the same temperature and volume emission measure.

\section{Conclusion}
\label{sec:Conclusion}

The use of column emission measures by CHIANTI and volume emission measures by routines such as OSPEX has caused some confusion when photon spectra generated by the two programs are compared for identical source conditions. We note that the relation between these two emission measures is just the area of the source as seen from the detector. However, for many X-ray observations, the source area is either unknown or imperfectly known for the spatially integrated spectra determined, e.g., from GOES observations. The units of the spectra produced by these two packages are different -- the spectral synthesis program ch\_ss in CHIANTI creates plots of the spectral radiance $I_E$ in units of photons cm$^{-2}$ s$^{-1}$ sr$^{-1}$ keV$^{-1}$ vs. photon energy in keV whereas OSPEX plots spectral irradiance $F_E$ in units photons cm$^{-2}$ s$^{-1}$ keV$^{-1}$, i.e., without the ``$sr^{-1}$.'' 

In this work, we derive the general relation between spectral radiance $I\lambda$ and spectral irradiance $F\lambda$ (Equation~\ref{eq:F_I_relation}) involving only the volume and column emission measures and the distance of the detector $D$ from the Sun. We illustrate the use of this relation by taking a radiance spectrum ($I_\lambda$) calculated with CHIANTI and transforming it into an irradiance spectrum vs. wavelength ($F_\lambda$) and vs. energy ($F_E$). 

We have developed a method for obtaining a spectral irradiance from CHIANTI for direct comparison with the corresponding spectral irradiance obtained with OSPEX. This method involves replacing the column emission measure input to CHIANTI with the same volume emission measure used by OSPEX but divided by the square of the distance between the source and the detector. This is essentially the same as the column emission measure times the unknown solid angle subtended by the source area at the detector. Hence, the spectrum generated by CHIANTI with this modified input is the spectral radiance times the solid angle, i.e., the same as the irradiance spectrum generated by OSPEX. At present, the CHIANTI spectral plot generated in this way is labeled with the ``$sr^{-1}$'' in the $y$-axis units. We suggest that a new feature be added to CHIANTI to allow the resulting spectrum to be labeled without the “$sr^{-1}$” in the units when the volume emission measure divided by $D^2$ is entered instead of the column emission measure.

This  method was used in generating the tables of spectra from the CHIANTI data base that are accessed by all the OSPEX thermal functions - vth, multitherm\_pow, multitherm\_exp, etc. Thus, the agreement between the spectra generated by the two packages shown in Figure~\ref{fig:spectra} is confirmation that the method was applied consistently in both directions. The small differences of up to 30\% may be caused by the different versions of CHIANTI that were used for the two cases - v8 for the OSPEX plot and v10.1 for the CHIANTI plot.

\section{Acknowledgments}

We acknowledge the very useful and now widely used CHIANTI atomic code and database, which is a collaborative project involving the University of Cambridge (UK), the NASA
Goddard Space Flight Center (USA), the George Mason University (GMU, USA) and the  University of Michigan (USA).

We thank Peter Young for help with CHIANTI and Kim Tolbert for help with the RHESSI data analysis software, particularly with OSPEX. We acknowledge the critical reading of an early draft by Albert Shih and Andrew Ingles and the valuable contribution of the referee in greatly improving this paper.

\appendix

\section{CHIANTI}
\label{sec:CHIANTI}

CHIANTI is a package of Interactive Data Language (IDL) procedures that uses an extensive data base to compute the expected line and continuum emission (free--free, free--bound, and two photon) from ionized astrophysical plasmas. The emission spectrum can be computed for an isothermal plasma or a plasma with a specified differential emission measure.  Temperatures ranging from $10^4$~K to 100~MK can be used with a selection of compositions, densities, and ionization states. 
 The latest version (10.1) used for this paper is described by \cite{2021ApJ...909...38D} and \cite{2023ApJS..268...52D}.  
 An IDL GUI is available (ch\_ss.pro) to make setting the input parameters, calculating a spectrum, and generating a plot or an output data file (.genx) relatively straightforward. The User Guide for CHIANTI Version 9 is available on Solar Software (SSW) at \url{/ssw/packages/chianti/doc/cug.pdf}. It gives details on using the GUI and on the individual IDL commands when not using the GUI. The main input parameters for an isothermal plasma are log$_{10}\,T$(K) and the column emission measure ($EM_h$) in units of cm$^{-5}$. 

Note that CHIANTI uses $N_H$ rather than $N_i$ in the calculation of column emission measures, where $N_H$ is the density of hydrogen. Since $N_H/N_i{\sim0.83}$ in a plasma with typical solar compositions, the column emission measures are smaller by this factor compared with those calculated using the density of all ions. While this approximate ratio is correct for high temperature sources observed with RHESSI where both H and He are completely ionized, it is not correct for sources with temperatures below $\sim$100,000~K where He, at least, is not completely ionized.

The generated spectrum of the spectral radiance ($I_E$) provided by CHIANTI is in units of photons~cm$^{-2}$~s$^{-1}$~sr$^{-1}$~keV$^{-1}$. Here, the $cm^{-2}$ refers to the area of the detector and the $sr^{-1}$ refers to the solid angle into which the photons are emitted (see Figure~\ref{fig:IntensityFlux}). All of the input and output data is stored in an IDL structure that can be saved in a .genx file and read back into an IDL procedure with the command, regen, for further customized display and analysis.

\section{Object Spectral Executive - OSPEX}
\label{sec:OSPEX}

OSPEX\footnote{OSPEX documentation is available on Solar Software (SSW) at \url{/ssw/packages/spex/doc/ospex_explanation.htm}} is an IDL object-oriented interface for X-ray spectral analysis of solar X-ray and gamma-ray observations \citep{2020ascl.soft07018T,2020ascl.soft07017S}. Typically, it uses as input the measured count-rate spectra from one of many different spectrometers and fits them to predicted photon spectra based on many different thermal and nonthermal models. It covers the photon energy range from 1.1 keV to 100 MeV and temperatures from $\sim5$ to $>100$ MK. 

Generally, limited spatial information is available in this energy range, and consequently the dimensions of the source are not generally known. Hence, OSPEX determines the spectral irradiance in units of photons~${\rm cm}^{-2}~{\rm s}^{-1}~{\rm keV}^{-1}$ rather than the spectral radiance provided by CHIANTI in units of photons ${\rm cm}^{-2}~{\rm s}^{-1}~{\rm keV}^{-1}~{\rm sr}^{-1}$, i.e., there is no ``per steradians'' in the flux units. The $cm^{-2}$ refers to the area of the detector as illustrated in Figure~\ref{fig:IntensityFlux}. Also, for thermal spectra, OSPEX provides the volume emission measure ($EM_V$) rather than the column emission measure ($EM_h$) used by CHIANTI.

\bibliography{EM}{}
\bibliographystyle{aasjournal}

\end{document}